\documentclass[a4paper, prl,amsmath,amssymb,floatfix,twocolumn]{revtex4}
\usepackage{graphicx}
\usepackage{color}
\usepackage{amsfonts}
\usepackage{amsmath}
\usepackage{times}
\def\E{{\bf E}}

\def\D{{\mathcal D}}

\def\A{{\bf A}}

\def\H{{\bf H}}

\def\G{{\bf G}}

\def\Curl{{\rm Curl}}

\def\om{\omega_0}

\def\rr{{\bf r}}

\begin{document}

\title{Numerical studies of Casimir interactions}

\author{S. Pasquali, A. C. Maggs} \affiliation{ Laboratoire de Physico-Chime
  Th\'eorique, Gulliver, CNRS-ESPCI, 10 rue Vauquelin, 75231 Paris Cedex 05,
  France.}
\begin{abstract}
  We study numerically the Casimir interaction between dielectrics in both two
  and three dimensions.  We demonstrate how sparse matrix factorizations
  enable one to study torsional interactions in three dimensions. In two
  dimensions we study the full cross-over between non-retarded and retarded
  interactions as a function of separation.  We use constrained factorizations
  in order to measure the interaction of a particle with a rough dielectric
  surface and compare with a scaling argument.
\end{abstract}

\maketitle

Dispersion forces have as their origin the fluctuations of polarization in
materials coupled to the long-ranged electrodynamic interaction described by
Maxwell's equations. The first convincing demonstration of their importance
was the calculation by Keesom of the interaction between fluctuating classical
dipoles \cite{keesom}.  The introduction of quantum fluctuations by London
\cite{london} accounted for the long-ranged, $1/r^6$, part of the van der
Waals interaction in most materials. Later, Casimir and Polder \cite{casimir}
showed that retardation modifies the interactions in an important manner--
leading to a decay in the interaction which is asymptotically $1/r^7$ at zero
temperature.  Further advances were made by the Russian school \cite{lifshitz}
who showed how to formulate the interactions in terms of the dielectric
response of materials.  Overviews with many references to theoretical and
experimental developments are to be found in \cite{ninham,review,Milton}.
Retarded Casimir interactions are the dominant interaction between neutral
surfaces at the submicron scale.

Whilst the analytic basis of the theory is largely established its application
is difficult in experimentally interesting geometries. One is constrained to
work with perturbative expansion about exactly solvable geometries
\cite{perturb}, or use {\it ad hoc} schemes such as the proximity force
approximation. Only a few geometries have been attacked with exact analytic
techniques \cite{emig}. Recently several attempts have been made to study
numerically the interactions by using methods from modern computational
science-- including fast multigrid lattice solvers \cite{mit} in order to
calculate Green functions and forces, or the use of discretized determinants
to determine free energies \cite{nitti}.

In this Letter we will present a series of techniques which enable one to
evaluate the interaction between dielectric bodies in full vectorial
electrodynamics.  Firstly, we calculate the torsional potential between two
three-dimensional bodies in the retarded regime, using a full discretization
of Maxwell's equations, we note that the Casimir torque has recently received
the attention of experimentalists \cite{Capasso,Shao}.  For more detailed
studies we present results for two-dimensional systems. This allows us to
study the cross-over between the near- and far-field regimes and also to
measure the interaction between a particle and a rough surface.  With these
two-dimensional systems we implement general strategies which substantially
increase the efficiency of simulations, at the same time decreasing the
sensitivity of the results to numerical round-off errors.

In three dimensions we discretize Maxwell's equations to a cubic Yee-lattice
\cite{yee}, lattice constant $a=1$, associating the electric degrees of
freedom to the links, magnetic degrees of freedom are localized on the faces
of the lattice. We remind the reader that the finite difference approximation
to the $\nabla \times$ operator, here designated $\Curl$, maps the electric
field on four links surrounding the face of the cube to the magnetic
field. $\Curl$ is needed in the Maxwell equation
\[
\frac{\partial \H }{ \partial t} = -c \;\Curl\, \E
\]
The adjoint operator maps fields from the faces to the links. We will denote
it $\Curl^*$. It intervenes in the second time dependent Maxwell equation.
\[
\frac {\partial \bf D }{ \partial t} = c\; \Curl^*\, \H
\]
The importance of clearly distinguishing the two operators will become
apparent when we discuss the two-dimensional case below.  We use
Heaviside-Lorentz units in which Maxwell's equations are directly
parameterized by the speed of light in vacuum, $c$.

From these two equations Lifshitz theory \cite{LanLif} shows that the free
energy of interaction between dielectric bodies is found from from the
imaginary time wave equation for the vector potential in the temporal gauge
where $\E=-\dot \A/c$ and $\phi=0$
\[
\left \{ \frac {\epsilon(\rr, \omega) \omega^2} { \hbar^2 c^2 } + \Curl^*
  \Curl \right \} {\A} = {\D}_A \, \A = 0
\]
Alternatively one introduces a magnetic formulation and works with a potential
such that $\H= \dot \G/c$ and considers the wave equation
\[
\left \{ \frac {\omega^2} { \hbar^2 c^2 } + \Curl \frac{1 }{ \epsilon(\rr,
    \omega)} \Curl^* \right \} {\G} = {\D}_G\, \G =0
\]

In our work we always consider the differences in free energy between pairs of
configurations; we thus avoid a full account of the self-energy variations of
dielectric media \cite{nitti}.  The free energy difference between two
configurations $1,2$ is found from
\begin{equation}
  U^{1,2} = \int_0^{\infty} \frac{d \omega}{2 \pi} 
  \left \{
    \ln  \det \,\D^1(\omega)-
    \ln \det \,\D^2(\omega)
  \right \}
  \label{F}
\end{equation}
for either choice of wave operator, $\D_A$ or $\D_G$; while self-energy
contributions are different in the two formulations we have verified with our
codes that both give the same result for the long-ranged part of the
interactions that we are interested in.

We perform the frequency integration in eq.~(\ref{F}) by changing integration
variables to $z$, where $ \omega= \alpha z/(1-z)$ with $0<z<1$. The parameter
$\alpha$ is chosen so that the major features in the integrand occur for
values of $z$ near $1/2$. We then use $N_g$-point Legendre-Gauss quadrature to
replace the integral by a weighted sum over discrete frequencies. We evaluate
determinants by finding the Cholesky factorization $L_D$ of $\D(\omega)$ such
that $L_D$ is lower triangular \cite{taucs} and $L_D L^T_D=\D(\omega)$. The
determinant of $\D$ is then given by
\[
\ln \det\D(\omega) = 2 \sum_i \ln \left( L_{D,i,i} \right )
\]

When we examine the detailed structure of Maxwell's equations discretized to
$V=L^3$ sites in three dimensions we discover that the $\Curl$ operator is a
matrix of dimension $3V \times 3V$ and has $12V$ non-zero elements.  The
operator $( \Curl^*\, \Curl)$ has $39V$ non-zero elements. The major technical
difficulty comes from the fact that the matrices we work with have dimensions
which are very large, $\sim 10^6 \times 10^6$. All numerical work was
performed with an Intel Xeon-5140 workstation.

\begin{figure}
  \includegraphics[scale=0.6]{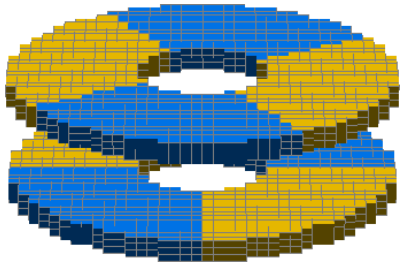}
  \caption{ A Pair of structured dielectric rings. Each quadrant has different
    dielectric properties.} \label{rings}
\end{figure}

We now calculate the Casimir torque between two parallel rings centered on a
common axis, figure ~\ref{rings}. Each ring is divided into quadrants with
alternating dielectric properties.  We take permittivities which are
independent of frequency, corresponding to the full retarded regime
\cite{LanLif} with $\epsilon_1(\omega)=5$, $\epsilon_2(\omega)=10$; the space
around the rings is a vacuum with $\epsilon_r=1$.  We measure the energy of
interaction as the top ring is rotated with respect to the lower. The zero of
the interaction corresponds to aligned rings.  As the rings are rotated the
interface between the dielectric materials, as interpolated to the lattice,
undergoes some re-arrangement changing the self energy of the rings. We thus
perform two runs. The first run of a single rotating ring determines this
variation in the self-energy. The second run with the both rings allows one to
measure the interaction energy by subtraction.

\begin{figure}[t]
  \includegraphics[scale=0.35]{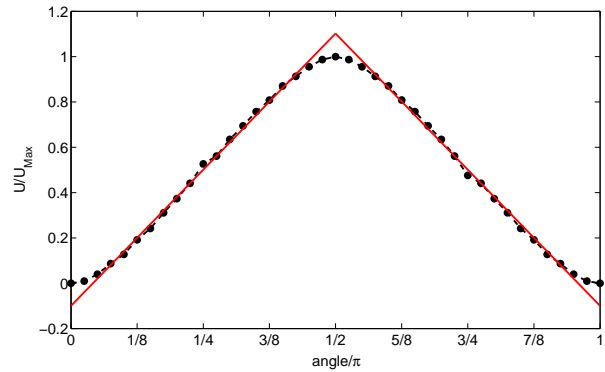}
  \caption{ Interaction energy as a function of angle as a ring of figure
    \ref{rings} rotates.  In the linear parts of the curve the torque is
    almost independent of the angle.  Ring diameter $36a$, separation and
    thickness $2a$.  Rounding is determined by the ratio of separation to
    diameter of the rings.  $10$ days of computation with $N_g=8$. Cholesky
    factor $9 {\rm GB}$.  } \label{ring}
\end{figure}

We worked with a system of dimensions $V=55 \times 55 \times 55$, figure
~\ref{ring}. The graph of the interaction energy as a function of angle is
noticeably triangular in shape between $\pi/8$ to $3\pi/8$. This is understood
by the fact that the interaction energy is dominated by the interactions
directly across the gap. The fluctuations in the curve about the expected
linear behavior, together with its slight asymmetry give an idea of the noise
coming from irregularities of the interpolation of the disks to the
lattice. This irregularity is particularly clear on comparing the points for
$\pi/4$ and $3 \pi/4$.

We now turn to two-dimensional electrodynamics where we can study systems with
larger linear dimensions. Such large system sizes are needed in order to
follow the cross-overs between different regimes in the interaction of
particles or if one wishes to simulate structured or disordered materials in
order understand the efficiency of analytic approximations.

In three dimensions the two formulation in terms of $\D_A$ and $\D_G$ are
largely equivalent. In two-dimensional electrodynamics this is no longer the
case. Consider an electrodynamic system in which there are two components of
the electric field in the $x-y$ plane; the magnetic field then has just a
single component in the $z$ direction. The $\Curl$ operator becomes a {\it
  rectangular matrix}\/ of dimensions $V \times 2V$ where now $V=L^2$. The
standard formulation in terms of the vector potential gives to an operator
$\D_A$ of dimensions $2V \times 2V$ with $14V$ non-zero elements; the
alternative formulation in terms of $\D_G$ leads to determinants of dimensions
$V \times V$ involving just $5V$ non-zero elements; the size of the matrix
that we must work with is smaller in the $\D_G$ formulation. We used $D_G$ in
the following numerical work, having checked that we obtain equivalent
results.

We started by measuring the cross-over between the short-ranged non-retarded
interaction to the long-ranged Casimir force. We studied a pair of dielectric
particles described by the single pole approximation to the dielectric
constant
\[
\epsilon(\omega)= 1 + \frac{\chi}{1 + \omega^2/\om ^2\hbar^2}
\]
where $\chi$ is the zero frequency electric susceptibility. The interaction is
retarded for separations $D \gg c/\om $, non-retarded for $D \ll c/\om$.

We measured the interaction between two dielectric particles in a square,
periodic cell of dimensions $L\times L$ using SuiteSparse \cite{davis} to
perform both the ordering and the factorization of the matrices.  We placed a
first particle at the origin, and considered two possible positions of a
second particle to calculate a free energy difference using eq.~(\ref{F}).
The first results were disappointing-- rather small systems ($L=50$) were
sensitive to numerical round-off errors. The origin of this problem was quite
clear. In a large system there is an extensive self-energy $\sim L^2$.  Pair
interactions calculated as the difference between two large numbers are
unreliable.

We avoided this problem by separating the free energy contributions from the
neighborhood of the three interesting sites and the rest of the system.  We
did this by introducing a block-wise factorization of $\D$ that enabled us to
both solve the round-off problem while re-using much of the numerical effort
need to generate the Cholesky factors thus improving the efficiency of the
code.

We now write the symmetric matrix from the wave equation in block form,
\begin{math} \D =
  \begin{pmatrix}
    X & \quad Y\\
    Y^T & \quad Z
  \end{pmatrix}
\end{math}.  Its determinant is $ \det(\D) = \det(X) \det(S)$ where the Schur
complement $S = Z - Y^T X^{-1} Y$ \cite{golub}. We group sites so that the
great majority is within the block $X$ and sites that we are interested in are
in the block $Z$. It is the term in $\det (X)$ the gives the large extensive
free energy which caused our numerical problems. It is independent of the
properties of our test particles. All the interesting information on energy
differences is in the Schur complement, $S$.

We start by finding the Cholesky factorization of $X$, $L_x$. The Schur
complement is calculated by solving the triangular equations $L_x U =Y$ by
forward substitution, then calculating $S=Z - U^T U$.  Our separation of
energies into an extensive constant and a small set of interacting sites
allows us to study the interaction of systems of sizes up to $L=2000$ before
round-off becomes a problem.

In order to generate data we generalized the method to a three level scheme--
firstly collect the set of sites (here $\sim 100$) of all the separations
required to generate a curve into the block $Z$, and form the Schur complement
forming a small effective theory for all these remaining sites.
Within the smaller matrix that has been generated we again re-order to
successively put each interesting sets of variables in the bottom-right corner
of the effective theory and find the Schur complement of these remaining
variables. We can then calculate interactions between the particles while
minimizing round-off errors.

\begin{figure}
  \includegraphics[scale=0.45]{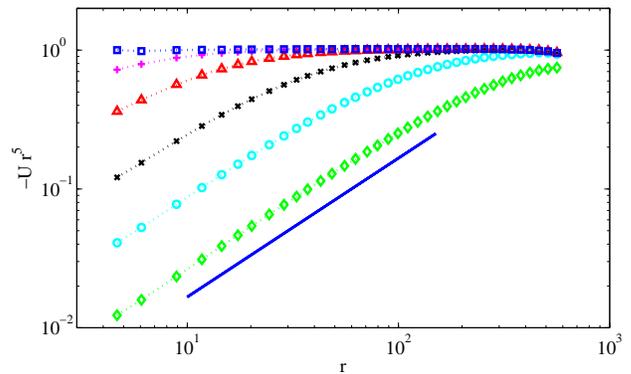}
  \caption{ Scaled interaction free energy, $-U r^5$ for a pair of dielectric
    particles ($\epsilon(0)=8$) in a box of dimensions $2000\times 2000$ as a
    function of separation. Curves from top to bottom correspond to $\om/c=$
    10, 0.3, 0.1, 0.03, 0.01 0.003. For large $\om/c$, $U r^5$ is constant,
    $\color{blue}\square$. For smaller $\om/c$ we see both retarded and
    non-retarded interactions.  Solid line corresponds to $U_{vdw}\sim
    1/r^4$. $10 {\rm GB}$ for Cholesky factor. Six hours of
    calculation. $N_g=25$.  } \label{U}
\end{figure}

We remind the reader that in two dimensions the electrostatic potential is
logarithmic between two charges, and that dipole-dipole fluctuations lead to
van der Waals interactions decaying as $U_{vdw}= 1/r^4$. As in three
dimensions retardation leads to an accelerated decay so that the Casimir
interaction varies as $U_c \sim 1/r^5$.  In our simulations we used values of
$\om/c$ varying from $0.003$ to $10$, figure \ref{U}. We determined the energy
of interaction of particles $U$, as a function of separation $r$ while moving
the second particle in the simulation cell out to $(L/5, L/5)$; the zero of
energy is calculated for two particles separated by $(L/2, L/2)$. We scale out
the retarded behavior, plotting $-U(r) r^5$. We see that for the largest
$\om/c$ the interactions are retarded for all separations,
$\color{blue}\square$.  For the smaller values of $\om/c$ the interaction
varies as $1/r^4$. In the scaled curve this gives the linear rise clearly
visible in the figure, $\color{green} \diamond$.  For $ 0.1 <\omega_0/c <0.01
$ we see both the near- and far-field behaviors clearly displayed within a
single sample-- permitting the detailed study of cross-over phenomena with
frequency dependent dielectric behavior. No assumptions of symmetry are made
in the calculation; the method can be used with bodies of arbitrary geometry.

\begin{figure}
  \includegraphics[scale=0.45]{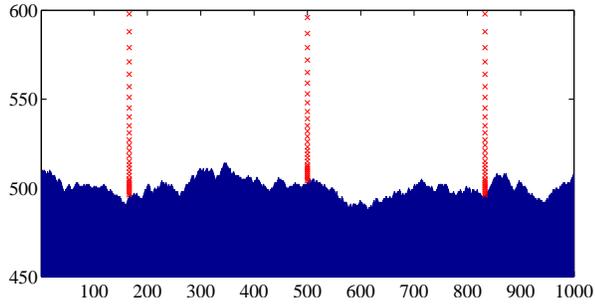}
\caption{ Realization of rough interface and set of measurement positions, ${\color{red} \times} $, for
  the interaction energy which will be separated into the block $Z$.
 Anisotropic horizontal and vertical scales.
\label{rough}
}
\end{figure}

We now turn to a problem where analytic results are much more difficult to
find: The interaction of a dielectric particle with a rough surface, figure
\ref{rough}. We generated rough surfaces as realizations of solid-on-solid
random walks on a lattice.  Approximately half of the simulation box contains
dielectric material with $\epsilon=8, \om=\infty$; the rest of the box has
$\epsilon=1$. We measure the interaction with a test particle as a function of
the distance from the rough surface using the above method of block Schur
complements to perform a single large factorization per frequency for each
realization of the disorder.  We generated $1000$ rough surfaces and measured
the average interaction with the surface $\langle U \rangle$, as a function of
separation, as well as the variance in the potentials.

\begin{figure}
  \includegraphics[scale=0.45]{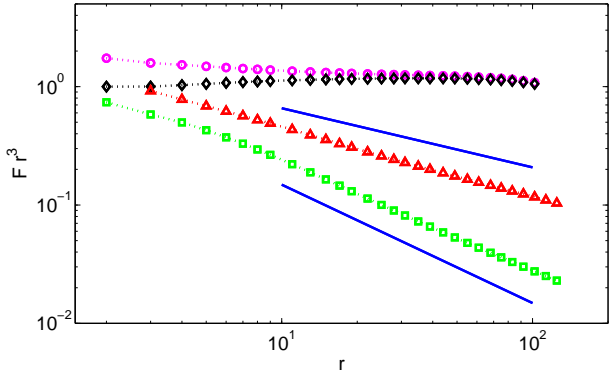}
  \caption{ (1) {\color {magenta} $\circ$}, $-\langle U \rangle r^3$, averaged
    interaction between dielectric particle and rough dielectric surface.  (2)
    $\diamond$, $-U_s r^3$, interaction between particle and flat surface.
    (3) {\color{red} $\triangle$}, $\sigma_u r^3$, variance of interaction for
    rough surfaces.  (4) {\color{green} $\square$}, $\delta U r^3$, difference
    in mean interaction energy between a flat and a rough surface.  Solid
    lines: $r^{-3.5}$ and $r^{-4}$. $L=1000$.  Two weeks of simulation
    time. Cholesky factor $2.5 {\rm GB}$. $N_g=20$.}
  \label{U2}
\end{figure}

We understand the results, figure \ref{U2}, with a scaling argument. When the
particle is a distance $r$ from the surface the interaction is dominated by a
front of length $r$ along the surface. Since the surface is a random walk its
average position is displaced by $\delta r \sim \pm r^{1/2}$ compared to the
flat surface.  The interaction between a smooth surface and a particle varies
as $U_s\sim 1/r^3$ in the Casimir regime.  The interaction of the particle
should thus be $U \sim 1/(r + \delta r)^3 $. If we expand to first order we
find that the variance of the interaction should scale as, ({\color{red}
  $\triangle$}) $\sigma_u \sim r^{-3.5}$ while the second order expansion
gives a shift in the mean potential, $\langle U\rangle$, which varies as,
(${\color{green} \square}$), $\delta U \sim 1/r^4$. The numerical data are
compatible with this scaling. The argument is easily generalized to affine
surfaces with other, less trivial roughness exponents giving results
compatible with \cite{li}.

We have demonstrated the power of direct methods from linear algebra when
applied to the study of dispersion forces.  In three dimensions we have
measured interactions in experimentally realizable geometries-- though system
sizes are still too small to accurately measure cross-overs between different
scaling regimes. In two dimensions we have shown how to measure the cross-over
between London dispersion and Casimir interactions, and have determined
correction to scaling exponents for the interactions of a disordered systems.

Work financed in part by Volkswagenstiftung.  \bibliography{mc}

\end{document}